\documentclass[12pt]{iopart}
\usepackage{iopams}
\usepackage{graphicx}
\usepackage{cite}
\usepackage{hyperref}
\usepackage{bm}

\expandafter\let\csname equation*\endcsname\relax
\expandafter\let\csname endequation*\endcsname\relax
\usepackage{amsmath}

\newcommand{\ket}[1]{\left|#1\right\rangle}

\usepackage{xcolor}
\usepackage[normalem]{ulem}
\usepackage{tikz} \usetikzlibrary{positioning}

\begin{document}

\title[Continuous-mode quantum optics of finite strong-field laser pulses]{A continuous-mode quantum-optical representation of finite strong-field laser pulses in free space}

\author{Szabolcs Hack$^{1,2,*}$, Attila Czirj\'{a}k$^{1,2}$ and P\'{e}ter F\"{o}ldi$^{1,2}$}

{$^1$ ELI ALPS, The Extreme Light Infrastructure ERIC, Wolfgang Sandner u. 3., 6728 Szeged, Hungary}%

{$^2$Department of Theoretical Physics, University of Szeged, Tisza Lajos k\"{o}r%
\'{u}t 84, H-6720 Szeged, Hungary}

{$^{*}$ corresponding author: szabolcs.hack@eli-alps.hu}

{quantum optics, ultrashort laser pulses, nonclassical light, strong-field physics, continuous modes}

\begin{abstract}
We formulate an energy-consistent quantum-optical representation of finite paraxial laser pulses in free space. The formulation is based on a continuous-mode description and provides a step toward a unified treatment of coherent and nonclassical strong-field drivers. Starting from plane-wave quantization, we construct a transverse vector-mode reduction in which the selected mode can
reproduce an arbitrary normalized spatial--polarization profile. For a coherent pulse, the measured pulse energy and full laboratory analytic signal determine the frequency-dependent coherent-state displacement $(\alpha(\omega))$, without introducing a
physical quantization volume. For nonclassical pulses, the same first-order field data and total energy do not uniquely specify the quantum state. Additional correlation
information or model assumptions are required. We incorporate this information through normal and anomalous covariance kernels and discuss $g^{(2)}$-based diagnostics of
squeezed pulses. We also derive a phase-sensitive convergence criterion for numerical frequency-bin discretizations. An exactly solvable continuum-mode free-electron
application shows that the coherent displacement reproduces the semiclassical mean trajectory, whereas the field covariances determine the electron wave packet width. The
framework offers a convenient, energy-consistent interface between measured free-space pulses and quantum-optical calculations.
\end{abstract}

\section{Introduction}

Finite-duration laser pulses are normally prescribed as classical fields in
strong-field and attosecond physics. This semiclassical description underlies
the three-step picture of high-harmonic generation (HHG), the strong-field
approximation, tunnel ionization, electron recollision, and attosecond-pulse
generation \cite{corkum_three_step,lewenstein_sfa,krausz_ivanov_attosecond}.
It is insufficient, however, when photon statistics or field correlations are
part of the physics.
Quantized-field descriptions of strong-field interactions, including HHG and its photon statistics, have been discussed for several decades and have attracted renewed attention in recent years \cite{bergou_varro1981,gonoskov_qosf,foldi,varro,gombkoto2016,gombkoto2020,gombkoto2024}. Photon-counting and correlation measurements have revealed quantum-optical signatures in the transmitted fundamental and harmonic fields \cite{tsatrafyllis2017,tsatrafyllis2019,theidel_qo_hhg}. Strong-field interactions have also been proposed as sources of high-photon-number optical cat states \cite{lewenstein2021_cat}, while HHG driven by incoherent classical and nonclassical light fields has been investigated in Ref. \cite{stammer_absence}.

The formal ingredients are well established: plane-wave continuum
quantization \cite{loudon,mandel_wolf}, continuous-mode coherent states
\cite{glauber,blow_loudon}, and wave packet modes for propagating light
\cite{sipe,smith_raymer}. Mode-basis freedom in multimode quantum optics is
reviewed in Ref.~\cite{fabre_treps}, while spectral-mode decompositions are
widely used to identify correlations in multiphoton states
\cite{rohde_spectral}. Related analytic treatments span nonlinear scattering
and photon--electron wave packet dynamics in quantized fields
\cite{varro2008_photon_electron,hack_wavepacket_2026}.
Despite these developments, a practical, energy-consistent framework connecting laboratory pulse information to a normalized free-space quantum state is usually left implicit. For coherent pulses, the relevant laboratory information consists of the spectral amplitude and phase of the electric field, which can be reconstructed using standard pulse-characterization techniques such as frequency-resolved optical gating (FROG) \cite{KaneTrebino1993} and spectral phase interferometry for direct electric-field reconstruction (SPIDER) \cite{IaconisWalmsley1998}. For nonclassical pulses, additional correlation information is required; this may be constrained through interferometric measurements of \(g^{(1)}\) and Hanbury Brown--Twiss measurements of \(g^{(2)}\) \cite{Glauber1963,HanburyBrownTwiss1956}. Making the correspondence between such measured pulse data and a normalized quantum state operational is a natural first step toward a unified pulse-level framework in which coherent and nonclassical strong-field drivers share a common mode normalization and consistent energy accounting, while their distinct photon statistics are encoded in the quantum state.

We construct that map in three steps. First, the plane-wave continuum is
rewritten in a complete transverse vector-mode basis. Any normalized spatial
and polarization profile, including a measured or distorted beam, can be chosen
as the first basis element; selecting only that element is the subsequent
physical single-mode approximation. Second, the remaining longitudinal
continuum is expressed in frequency-normalized operators. Third, the coherent
amplitude \(\alpha(\omega)\) is chosen so that the quantum mean field equals the
prescribed classical analytic signal. In this representation, the individuality of the pulse is encoded in its quantum state rather than in a pulse-dependent operator.

Matching the first moment alone leaves the quantum state underdetermined. For
coherent pulses, the coherent-state assumption completes the map. For displaced
squeezed pulses, the additional second-order information is encoded in normal
and anomalous covariance kernels, allowing states with the same mean waveform
to have different fluctuations and $g^{(2)}$ signatures. We discuss
mode-resolved and broadband diagnostics and derive a phase-sensitive
convergence criterion for frequency-bin calculations. Finally, an exactly
solvable continuum-mode free-electron model shows that the displacement
determines the mean trajectory, whereas the covariances determine the
wave packet width.

The paper is organized as follows. Section~\ref{sec:field_quantization} fixes
the continuum convention, selects and normalizes the beam mode.
Section~\ref{sec:coherent} develops the coherent-state map,
its laboratory implementation, and the frequency-bin convergence criterion. Section~\ref{sec:squeezed} extends the description to displaced squeezed pulses, discusses second-order diagnostics, and applies the resulting continuum-state specification to an exactly solvable free-electron model. Section~\ref{sec:relations} relates the construction to other
formalisms. Technical derivations are collected in the appendices.

\section{Free-space quantization and transverse-mode reduction}
\label{sec:field_quantization}

\subsection{Plane-wave continuum and normalization}

In the Coulomb gauge, the positive-frequency electric field of the source-free
plane-wave continuum is
\begin{equation}
\hat{\mathbf E}^{(+)}(\mathbf r,t)
=
i\sum_{\lambda=1,2}
\int \frac{d^3k}{(2\pi)^{3/2}}
\sqrt{\frac{\hbar\omega_k}{2\varepsilon_0}}\,
\bm{\epsilon}_{\mathbf k\lambda}
e^{i\mathbf k\cdot\mathbf r}
\hat a(\mathbf k,\lambda)e^{-i\omega_k t},
\label{eq:E_plane}
\end{equation}
where \(\omega_k=c|\mathbf k|\), the vectors
\(\bm{\epsilon}_{\mathbf k\lambda}\) form an orthonormal transverse
polarization basis, and
\begin{equation}
[\hat a(\mathbf k,\lambda),\hat a^\dagger(\mathbf k',\lambda')]
=
\delta_{\lambda\lambda'}\delta^{(3)}(\mathbf k-\mathbf k').
\label{eq:plane_comm}
\end{equation}
Eq.~\eref{eq:E_plane} follows either directly or as the continuum limit of
box quantization \cite{loudon,mandel_wolf,blow_loudon}. The auxiliary volume
cancels after the operators are rescaled to obey Eq.~\eref{eq:plane_comm}; the
derivation is recalled in \ref{app:V}. Relative to a chosen propagation
axis \(z\), we write
\begin{equation}
\mathbf k=(\mathbf k_\perp,k_z),\qquad d^3k=d^2k_\perp\,dk_z.
\end{equation}
This relabeling of the axes merely anticipates the fact that a realistic laser pulse occupies a narrow angular distribution around a chosen direction and is naturally
associated with a restricted family of transverse modes adapted to the beam geometry.

\subsection{Transverse-mode basis and projection}

For each fixed value of \(k_z\), let the coefficient functions \(\widetilde u_{m\lambda}(\mathbf k_\perp;k_z)\) form a complete orthonormal set in the combined transverse-wave-vector and polarization space. These functions are the Fourier transforms of the corresponding real-space vector modes. Here \(m\) labels the transverse mode, whereas \(\lambda\) labels its components in the local plane-wave polarization basis \(\bm\epsilon_{\mathbf k\lambda}\), with \(\bm\epsilon_{\mathbf k\lambda}\cdot\mathbf k=0\).
 We assume
\begin{equation}
\sum_{\lambda=1,2}
\int d^2k_\perp\,
\tilde u^*_{m\lambda}(\mathbf k_\perp;k_z)
\tilde u_{m'\lambda}(\mathbf k_\perp;k_z)
=
\delta_{mm'} .
\label{eq:transverse_vector_orth}
\end{equation}
The corresponding annihilation operators are defined by
\begin{equation}
\hat a_m(k_z)
=
\sum_{\lambda=1,2}
\int d^2k_\perp\,
\tilde u^*_{m\lambda}(\mathbf k_\perp;k_z)
\hat a(\mathbf k_\perp,k_z,\lambda).
\label{eq:am_def}
\end{equation}
Using the plane-wave commutation relations, one obtains
\begin{equation}
[\hat a_m(k_z),\hat a^\dagger_{m'}(k_z')]
=
\delta_{mm'}\delta(k_z-k_z').
\label{eq:am_comm}
\end{equation}
The commutator and the associated completeness relation are derived in \ref{app:mode_transform}.

The first mode need not be Gaussian, Hermite--Gaussian, or any other named
solution. For each \(k_z\), one may choose any normalized angular spectrum, whether measured or calculated, as \(\tilde u_{0\lambda}\), including aberrations or spatial--spectral coupling, and complete it by Gram--Schmidt orthogonalization.
All modes must be retained for an exact basis change; keeping only \(m=0\) is
accurate when the orthogonal modes are in vacuum or irrelevant to the
measurement. If a frequency-independent transverse basis is required, a
strongly space--frequency-coupled pulse may instead occupy several values of
\(m\).

\subsection{Single-mode paraxial continuum}
\label{sec:paraxial}

We now restrict attention to one selected transverse mode, denoted by \(m=0\),
and write
\begin{equation}
\hat a(k_z)\equiv \hat a_0(k_z).
\end{equation}
For a forward paraxial beam, $|\mathbf k_\perp|\ll k_z$. This approximation has two important consequences. First, the residual transverse-wave-vector dependence of the
dispersion relation may be neglected,
\begin{equation}
\omega_k=c\sqrt{k_z^2+|\mathbf k_\perp|^2}\simeq ck_z ,
\end{equation}
with a relative error of order $\langle k_\perp^2\rangle/k_z^2$. Second, the exact plane-wave transversality condition
$\mathbf k\cdot\bm\epsilon_{\mathbf k\lambda}=0$ implies that the longitudinal field component is of relative order
$|\mathbf k_\perp|/k_z$. This component is neglected at the paraxial order retained here, so that the real-space vector profile introduced below may be taken to satisfy
$\mathbf e_z\cdot\mathbf u_\omega=0$.
We introduce frequency-normalized operators by
\begin{equation}
\hat a(\omega)
=
\sqrt{\frac{dk_z}{d\omega}}\,
\hat a(k_z).
\label{eq:a_frequency_definition}
\end{equation}
The square-root Jacobian accounts for the change of continuum variable and
preserves the canonical delta normalization, so that
\begin{equation}
[\hat a(\omega),\hat a^\dagger(\omega')]
=
\delta(\omega-\omega').
\label{eq:aw_comm}
\end{equation}
The field operator then becomes
\begin{equation}
\hat{\mathbf E}^{(+)}(\mathbf r,t)
=
i
\int_0^\infty d\omega\,
\mathcal G(\mathbf r,\omega)\,
\hat a(\omega)
e^{-i\omega t}.
\label{eq:E_continuum_mode}
\end{equation}
All spatial and polarization factors are contained in the vector
mode function $\mathcal G$. Transforming the angular-spectrum modes to real space, the normalization in
Eq.~\eref{eq:transverse_vector_orth} implies that the corresponding real-space
vector mode has unit transverse norm. We write this mode as
$\mathbf u_\omega(\mathbf r_\perp,z)/\sqrt{A_{\rm m}}$, where
$A_{\rm m}$ has dimensions of area and $\mathbf u_\omega$ is
dimensionless. For positive-frequency quantities, we adopt the temporal
Fourier convention
\begin{equation}
F(t)=\int_{0}^{\infty}d\omega\widetilde F(\omega)e^{-i\omega t},
\qquad
\widetilde F(\omega)=\frac{1}{2\pi}\int_{-\infty}^{\infty}dt F(t)e^{i\omega t}.
\label{eq:fourier_convention}
\end{equation}
The field operator can then be written as
\begin{equation}
\hat{\mathbf E}^{(+)}(\mathbf r,t)
=
 i
\int_0^\infty d\omega\,
\sqrt{\frac{\hbar\omega}{4\pi\varepsilon_0 c A_{\rm m}}}
\,\mathbf u_\omega(\mathbf r_\perp,z)\,
\hat a(\omega)e^{-i\omega t}.
\label{eq:paraxial_G_convention}
\end{equation}
Accordingly, the dimensionless profile satisfies
\begin{equation}
\int d^2r_\perp\,
\mathbf u_\omega^*(\mathbf r_\perp,z)\!\cdot\!
\mathbf u_\omega(\mathbf r_\perp,z)=A_{\rm m}.
\label{eq:transverse_realspace_norm}
\end{equation}
Thus
\(\mathcal G(\mathbf r,\omega)=
\sqrt{\hbar\omega/(4\pi\varepsilon_0 c A_{\rm m})}\,
\mathbf u_\omega\), where the area \(A_{\rm m}\) fixes a transverse
normalization convention, not a quantization area. The transverse-mode transformation
including the Jacobian in Eq.~\eref{eq:a_frequency_definition} and the paraxial
reduction leading to Eq.~(\ref{eq:paraxial_G_convention}) are derived in \ref{app:mode_transform}.

Within this paraxial description, the vector profile $\mathbf u_\omega$
can represent arbitrary linear, elliptical, or frequency-dependent pure
polarization. Clearly, if more than one polarization mode is relevant, the corresponding
mode indices and operators must be retained.

\section{Multimode coherent states as laser pulses}
\label{sec:coherent}

\subsection{Continuous-mode coherent states}
Once the transverse structure has been fixed, the remaining longitudinal
continuum may be labeled by frequency. A classical laser pulse
occupying this mode family is represented quantum mechanically by a
continuous-mode coherent state,
\begin{equation}
\ket{\{\alpha\}}
=
\hat D[\alpha]\ket{0},
\qquad
\hat D[\alpha]
=
\exp\!\left[
\int_0^\infty d\omega\,
\left(
\alpha(\omega)\hat a^\dagger(\omega)
-
\alpha^*(\omega)\hat a(\omega)
\right)
\right].
\label{eq:coh_cont}
\end{equation}
Here \(\alpha(\omega)\) is a square-integrable spectral amplitude. The state is
an eigenstate of the annihilation operators,
\begin{equation}
\hat a(\omega)\ket{\{\alpha\}}
=
\alpha(\omega)\ket{\{\alpha\}},
\label{eq:a_eigen}
\end{equation}
and therefore
\begin{equation}
\left\langle \{\alpha\} \right|
\hat{a}(\omega)
\left| \{\alpha\} \right\rangle
=
\alpha(\omega).
\label{eq:a_mean}
\end{equation}
Its mean photon number and free-field energy above the vacuum are
\begin{equation}
\bar N
=
\int_0^\infty d\omega\,|\alpha(\omega)|^2,
\label{eq:N_alpha}
\end{equation}
and
\begin{equation}
U
=
\int_0^\infty d\omega\,
\hbar\omega\,|\alpha(\omega)|^2,
\label{eq:energy_alpha}
\end{equation}
respectively.

This construction follows the standard continuous-mode coherent-state formalism
\cite{glauber,blow_loudon}. It is also consistent with the
general modal viewpoint that a pure multimode coherent state can be regarded,
in an adapted mode basis, as a coherent excitation of a single effective mode
with the remaining modes in vacuum \cite{fabre_treps}. Its role here is to provide the
state-space representation onto which an experimentally specified laser pulse
will be mapped.

\subsection{Matching the mean field: $\alpha(\omega)$}

Let the classical positive-frequency electric field in the selected mode family
be written as
\begin{equation}
\mathbf E^{(+)}_{\mathrm{cl}}(\mathbf r,t)
=
\int_0^\infty d\omega\,
\bm{\mathcal E}_{\mathrm{cl}}(\mathbf r,\omega)e^{-i\omega t}.
\label{eq:Ecl_spectral}
\end{equation}
The Fourier convention is Eq.~\eref{eq:fourier_convention}. In the coherent
state, the expectation value of the field operator in
Eq.~\eref{eq:E_continuum_mode} is
\begin{equation}
\langle \hat{\mathbf E}^{(+)}(\mathbf r,t)\rangle
=
i
\int_0^\infty d\omega\,
\mathcal G(\mathbf r,\omega)\alpha(\omega)e^{-i\omega t}.
\label{eq:E_mean}
\end{equation}
If the prescribed classical field belongs to the selected mode family, then its
spectral amplitude factorizes as
\begin{equation}
\bm{\mathcal E}_{\mathrm{cl}}(\mathbf r,\omega)
=
i\,\mathcal G(\mathbf r,\omega)\alpha(\omega).
\label{eq:mode_matching}
\end{equation}
Equivalently, the coherent amplitude is fixed by the mode-matching condition
\begin{equation}
\langle \hat{\mathbf E}^{(+)}(\mathbf r,t)\rangle
=
\mathbf E^{(+)}_{\mathrm{cl}}(\mathbf r,t).
\label{eq:matching_condition}
\end{equation}
For an interaction that samples the electric field locally, the same condition may be imposed at the interaction point \(\mathbf r_0\),
\begin{equation}
\alpha(\omega)
=
-i\,
\frac{\mathcal G^*(\mathbf r_0,\omega)\!\cdot\!
\bm{\mathcal E}_{\mathrm{cl}}(\mathbf r_0,\omega)}
{\mathcal G^*(\mathbf r_0,\omega)\!\cdot\!
\mathcal G(\mathbf r_0,\omega)},
\label{eq:alpha_match}
\end{equation}
provided that the classical field is parallel to the selected vector mode at
\(\mathbf r_0\). The expression is otherwise the least-squares projection onto
that mode; the residual belongs to orthogonal spatial or polarization modes. 

Once the mode family has been fixed, the temporal envelope is encoded in the coherent amplitude \(\alpha(\omega)\), whose Fourier synthesis reconstructs the prescribed classical field.
Importantly, Eq.~\eref{eq:matching_condition} fixes only the first field moment and specifies a unique quantum state only after the coherent-state assumption is made. Section ~\ref{sec:squeezed} extends the construction to nonclassical states that may share the same mean field while differing in their covariance and correlation properties. 

If desired, one may decompose the coherent amplitude into an overall amplitude
and a normalized spectral shape,
\begin{equation}
\alpha(\omega)
=
\alpha_0\,\xi(\omega),
\qquad
\int_0^\infty d\omega\,|\xi(\omega)|^2=1.
\label{eq:alpha_xi}
\end{equation}
Then
\begin{equation}
\bar N=|\alpha_0|^2,
\qquad
U=|\alpha_0|^2
\int_0^\infty d\omega\,\hbar\omega|\xi(\omega)|^2.
\label{eq:energy_xi}
\end{equation}
This form leads directly to the wave packet mode description discussed in
Sec.~\ref{sec:relations}; however, introducing a separate wave packet operator
is a matter of convenience, not a change of the underlying field operator.

\subsection{General construction for a prescribed transverse and temporal pulse}
\label{sec:recipe}

The preceding subsection maps a prescribed classical field onto a continuous-mode coherent state. We now use this result to specify an input classical field, convert it into the spectral amplitude appearing in Eq.~(\ref{eq:Ecl_spectral}), and obtain the coherent amplitude from Eq.~(\ref{eq:mode_matching}) for an exactly mode-matched field, from Eq. (\ref{eq:projected_alpha_recipe}) by transverse projection, or from Eq. ~(\ref{eq:alpha_match}) for a local interaction.

The laboratory data determine the vector analytic signal in
Eq.~\eref{eq:Ecl_spectral}. Its spectrum can be written schematically as
\begin{equation}
\bm{\mathcal E}_{\rm cl}(\mathbf r,\omega)
=
|\mathcal E_{\rm cl}(\mathbf r,\omega)|\,
\mathbf e(\mathbf r,\omega)e^{i\Phi_{\rm cl}(\mathbf r,\omega)},
\qquad
\mathbf e^*\!\cdot\!\mathbf e=1,
\label{eq:classical_spectral_decomposition}
\end{equation}
where, after fixing its overall phase convention, the transverse complex unit
vector \(\mathbf e\) contains arbitrary pure polarization.  Eq.~\eref{eq:mode_matching} transfers it to
\(\alpha(\omega)\) relative to the phase convention of \(\mathcal G\); a
frequency-dependent rephasing of the operator and mode function redistributes
these phases but leaves the reconstructed field invariant. Around the carrier,
the classical spectral phase may be expanded as
\begin{equation}
\Phi_{\rm cl}(\omega)=\varphi_{\rm CEP}+\Phi_1(\omega-\omega_0)
+\frac{\Phi_2}{2}(\omega-\omega_0)^2+\cdots,
\label{eq:spectral_phase_expansion}
\end{equation}
where \(\Phi_1\) shifts the time origin, \(\Phi_2\) is the group-delay
dispersion that produces linear chirp, and higher orders describe more general
phase shaping \cite{weiner_ultrafast}. Spatially dependent phase and polarization may instead be
included in the frequency-dependent vector mode \(\mathbf u_\omega\).

For the simpler case of a separable, linearly polarized paraxial pulse at its
waist, one may write
\begin{equation}
\mathbf E^{(+)}_{\rm cl}(\mathbf r_\perp,t)
=
\mathbf e_x
\frac{E_{0}}{2}\,
F(\mathbf r_\perp)\,f(t)\,
\exp[-i\omega_0 t+i\varphi_{\rm CEP}+i\phi(t)],
\label{eq:lab_field_general}
\end{equation}
where \(F\) and \(f\) are dimensionless field envelopes and \(\phi(t)\) is an
equivalent time-domain description of phase modulation. The real field is
\(\mathbf E_{\rm cl}=2\operatorname{Re}\mathbf E_{\rm cl}^{(+)}\), and its
cycle-averaged paraxial intensity is
\(I=(\varepsilon_0 c/2)E_0^2|F|^2|f|^2\). The peak amplitude \(E_{0}\) is
fixed by the measured pulse energy. In the usual paraxial energy-flux normalization,
\begin{equation}
U
=
\frac{\varepsilon_0 c}{2}E_{0}^2
\left[\int d^2r_\perp\,|F(\mathbf r_\perp)|^2\right]
\left[\int_{-\infty}^{\infty}dt\,|f(t)|^2\right],
\label{eq:general_energy_normalization}
\end{equation}
when the peak field envelope is normalized to unity.

The spectral amplitude is obtained using the
Fourier convention in Eq.~\eref{eq:fourier_convention}. For a mode function
of the form
\begin{equation}
\mathcal G(\mathbf r,\omega)
=
G_0(\omega)\mathbf u_\omega(\mathbf r_\perp,z),
\end{equation}
the coherent amplitude associated with the selected mode is obtained by
transverse projection. Using the normalization in
Eq.~\eref{eq:transverse_realspace_norm}, one finds
\begin{equation}
\alpha(\omega)
=
-\frac{i}{G_0(\omega)A_{\rm m}}
\int d^2r_\perp\,
\mathbf u_\omega^*(\mathbf r_\perp,z)\cdot
\boldsymbol{\mathcal E}_{\rm cl}(\mathbf r,\omega).
\label{eq:projected_alpha_recipe}
\end{equation}

A field is exactly mode--matched when, at every frequency, its transverse
spatial--polarization profile is proportional to
\(\mathbf u_\omega\), with no component in any orthogonal transverse mode.
Within the present reference-plane formulation, the projection in
Eq.~\eref{eq:projected_alpha_recipe} is evaluated in the plane in which the
transverse mode is specified. For such an exactly mode-matched field
interacting with a localized dipole, one may equivalently use the local
expression in Eq.~\eref{eq:alpha_match}, provided that the transverse mode,
and hence the energy normalization, has already been fixed. The workflow for constructing a continuous-mode coherent state from laboratory pulse data is visualized in Fig.~1.

\begin{figure}[t]
\centering 
\includegraphics[width=\linewidth]{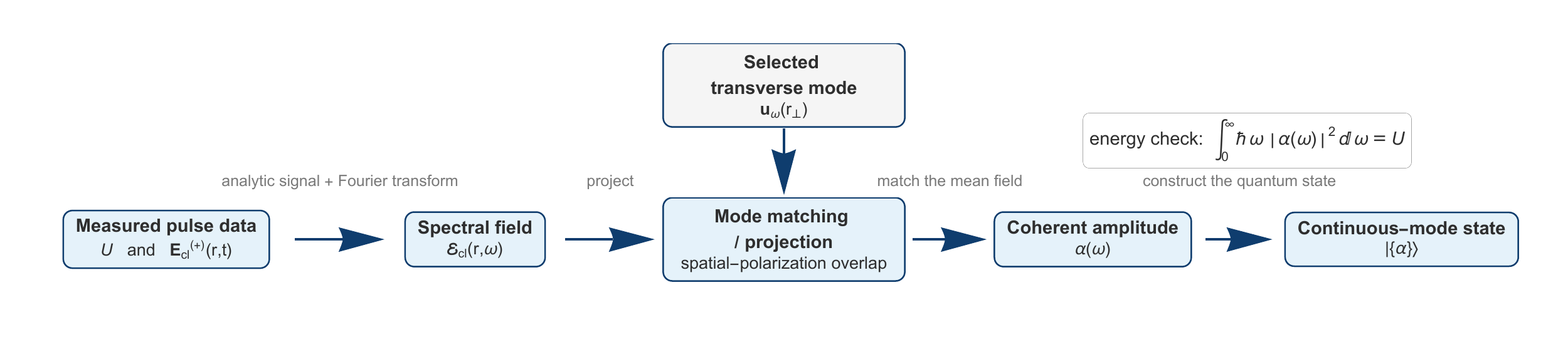} 
\centering
\caption{
Workflow for constructing a continuous-mode coherent state from laboratory pulse data. The measured pulse determines the analytic signal and its spectrum, while the chosen
transverse mode family defines the spatial-polarization degree of freedom. Mode matching (or projection onto the selected mode family) yields the coherent amplitude
$\alpha(\omega)$, whose normalization is verified by reproducing the measured pulse energy before constructing the coherent state $\ket{\{\alpha\}}$. }
\label{fig:coherent_state_workflow} \end{figure}

The consistency checks are those already given in Eqs.~\eref{eq:N_alpha} and
\eref{eq:energy_alpha}. In particular, the energy computed from
\(\alpha(\omega)\) must reproduce the measured pulse energy. For a narrowband
pulse centered at \(\omega_0\), this reduces to the useful estimate
\begin{equation}
\bar N\simeq \frac{U}{\hbar\omega_0},
\label{eq:narrowband_photon_number_recipe}
\end{equation}
but the exact definition of \(\bar N\) remains Eq.~\eref{eq:N_alpha}. If desired,
one may finally introduce the normalized pulse mode \(\xi(\omega)\) of
Eq.~\eref{eq:alpha_xi}; then the laboratory construction supplies the particular
\(\xi(\omega)\) and the amplitude \(\alpha_0\).

\subsection{Gaussian spatiotemporal pulse: numerical example}
\label{sec:numerical_example}

We now apply the preceding prescription to a transform-limited Gaussian pulse.
The transverse intensity profile at the waist is taken to be
\begin{equation}
I(r_\perp,t)=I_{0}
\exp\!\left(-\frac{2r_\perp^2}{w_0^2}\right)
\exp\!\left[-4\ln2\,\frac{t^2}{\tau_{\rm FWHM}^2}\right],
\label{eq:gaussian_intensity_example}
\end{equation}
so that the effective power area is
\begin{equation}
A_{\rm pow}=\int d^2r_\perp\,
\exp\!\left(-\frac{2r_\perp^2}{w_0^2}\right)
=\frac{\pi w_0^2}{2}.
\label{eq:gaussian_area_example}
\end{equation}
For the dimensionless field mode
\(\mathbf u=\mathbf e_x\exp(-r_\perp^2/w_0^2)\), this is also the
normalization area \(A_{\rm m}\) in Eq.~\eref{eq:transverse_realspace_norm}.
The energy normalization gives
\begin{equation}
I_{0}
=
\frac{U}{A_{\rm pow}\tau_{\rm FWHM}}
\sqrt{\frac{4\ln2}{\pi}},
\label{eq:example_Ipk}
\end{equation}
and therefore
\begin{equation}
E_{0}
=
\sqrt{\frac{2I_{0}}{c\varepsilon_0}}.
\label{eq:example_Epk}
\end{equation}
Equations~\eref{eq:example_Ipk} and \eref{eq:example_Epk} fix the classical
field scale entering Eq.~\eref{eq:lab_field_general}. The corresponding coherent
amplitude is obtained by inserting the positive-frequency spectrum of this
Gaussian pulse into Eq.~\eref{eq:mode_matching} or Eq.~\eref{eq:alpha_match}.

For the representative parameters listed in Table~\ref{tab:example_pulse}, Eqs. (\ref{eq:energy_xi}), (\ref{eq:example_Ipk}), and (\ref{eq:example_Epk}) determine the mean photon number, the pulse-mode coherent amplitude, the peak intensity, and the peak electric field. For a transform-limited Gaussian pulse, \(\Delta\nu_{\mathrm{FWHM}}\tau_{\mathrm{FWHM}}\simeq0.44\), which fixes the spectral bandwidth. Within the narrowband approximation, a convenient normalized photon-mode amplitude is therefore
\begin{equation}
\xi(\omega)
=
\left(\frac{1}{2\pi\sigma_\omega^2}\right)^{1/4}
\exp\!\left[-\frac{(\omega-\omega_0)^2}{4\sigma_\omega^2}\right],
\qquad
\sigma_\omega=
\frac{\Delta\omega_{\rm FWHM}}{2\sqrt{2\ln2}},
\label{eq:example_xi}
\end{equation}
with \(\int_0^\infty d\omega\,|\xi(\omega)|^2\simeq1\). In the notation of
Eq.~\eref{eq:alpha_xi}, the coherent amplitude can be written as
\begin{equation}
\alpha(\omega)=\sqrt{\bar N}\,\xi(\omega)e^{i\Phi_\alpha(\omega)},
\label{eq:example_alpha_omega}
\end{equation}
where
\(\Phi_\alpha=\Phi_{\rm cl}-\arg[iG_0(\omega)]\) contains the
carrier-envelope phase, chirp, and any higher-order spectral phase relative to
the selected mode convention. Substitution into Eq.~\eref{eq:coh_cont} gives
the quantum state associated with the specified laboratory pulse. The
agreement between the prescribed laboratory pulse energy and the coherent-state
energy calculated from Eq.~\eref{eq:energy_alpha} provides the required
energy-consistency check.

\begin{table}[t]
\centering
\caption{Quantum-optical parameters of a representative Gaussian femtosecond
pulse with \(\lambda_0=1030\,{\rm nm}\), \(\tau_{\rm FWHM}=330\,{\rm fs}\),
\(U=0.1\,{\rm mJ}\), and \(w_0=10\,\mu{\rm m}\).}
\begin{tabular}{lll}
\hline
Quantity & Symbol & Value \\
\hline
Central wavelength & \(\lambda_0\) & \(1030\,{\rm nm}\) \\
Central angular frequency & \(\omega_0\) & \(1.83\times10^{15}\,{\rm s}^{-1}\) \\
Photon energy & \(\hbar\omega_0\) & \(1.93\times10^{-19}\,{\rm J}\) \\
Pulse energy & \(U\) & \(0.1\,{\rm mJ}\) \\
Mean photon number & \(\bar N\) & \(5.2\times10^{14}\) \\
Mode coherent amplitude & \(|\alpha_0|\) & \(2.3\times10^7\) \\
Intensity FWHM duration & \(\tau_{\rm FWHM}\) & \(330\,{\rm fs}\) \\
Spectral FWHM & \(\Delta\nu_{\rm FWHM}\) & \(1.33\,{\rm THz}\) \\
Relative bandwidth & \(\Delta\omega_{\rm FWHM}/\omega_0\) & \(4.6\times10^{-3}\) \\
Beam waist & \(w_0\) & \(10\,\mu{\rm m}\) \\
Peak intensity & \(I_{0}\) & \(1.8\times10^{14}\,{\rm W/cm^2}\) \\
Peak electric field & \(E_{0}\) & \(3.7\times10^10\,{\rm V/m}\) \\
\hline
\end{tabular}
\label{tab:example_pulse}
\end{table}

\subsection{Frequency-bin discretization and convergence}
\label{sec:freq_bins}

For numerical work, partition the spectrum into intervals \(B_n\) of width
\(\Delta\omega\) and define
\begin{equation}
\hat a_n=\frac{1}{\sqrt{\Delta\omega}}
\int_{B_n}d\omega\,\hat a(\omega),
\qquad
\alpha_n=\frac{1}{\sqrt{\Delta\omega}}
\int_{B_n}d\omega\,\alpha(\omega).
\label{eq:bin_definitions_main}
\end{equation}
These operators obey \([\hat a_n,\hat a_m^\dagger]=\delta_{nm}\). With
\(\bar\omega_n\) denoting the midpoint of bin \(B_n\), the discrete photon
number and energy are
\begin{equation}
N_{\Delta\omega}=\sum_n|\alpha_n|^2,
\qquad
U_{\Delta\omega}=\sum_n\hbar\bar\omega_n|\alpha_n|^2.
\label{eq:bin_invariants_main}
\end{equation}
These quantities converge to Eqs.~\eref{eq:N_alpha} and \eref{eq:energy_alpha} as
\(\Delta\omega\to0\) and the spectral window is enlarged until the omitted
tails are negligible. At finite resolution, \(N_{\Delta\omega}\) is the photon
number carried by the piecewise-constant projection of \(\alpha(\omega)\), not
the exact photon number integrated over that window. Its phase dependence thus
measures unresolved intrabin structure rather than physical photon loss.

For a smooth spectrum and negligible truncation error,
 \ref{app:frequency_bins} gives
\begin{equation}
\epsilon_N
\equiv 1-\frac{N_{\Delta\omega}}{\bar N}
=
\frac{(\Delta\omega)^2}{12}
\frac{\displaystyle\int d\omega\,
|\partial_\omega\alpha(\omega)|^2}
{\displaystyle\int d\omega\,|\alpha(\omega)|^2}
+O\!\left((\Delta\omega)^4\right).
\label{eq:bin_asymptotic_main}
\end{equation}
Thus the dimensionless global resolution parameter is
\(\Delta\omega\sqrt{R_\alpha}\), where
\(R_\alpha=\int|\partial_\omega\alpha|^2d\omega/
\int|\alpha|^2d\omega\). It contains both amplitude and phase gradients. For
the Gaussian mode in Eq.~\eref{eq:example_xi}, with
\(\Phi_\alpha=C(\omega-\omega_0)^2/(2\sigma_\omega^2)\),
\begin{equation}
\epsilon_N
=\frac{1+4C^2}{48}
\left(\frac{\Delta\omega}{\sigma_\omega}\right)^2
+O\!\left[\left(\frac{\Delta\omega}{\sigma_\omega}\right)^4\right].
\label{eq:gaussian_bin_asymptotic_main}
\end{equation}
The numerical test in \ref{app:frequency_bins} verifies this
second-order convergence for transform-limited and chirped pulses. Through the
mode-matching relation \(\bm{\mathcal E}_{\rm cl}=i\mathcal G\alpha\), the same
criterion controls the discretization of the classical field and its coherent
state; no separate classical and quantum bin densities are required.
Convergence of the reconstructed field and the observables of interest should
nevertheless be checked explicitly.

\section{Nonclassical pulses and the response of matter}
\label{sec:squeezed}

The coherent-state construction of Sec.~\ref{sec:coherent} fixes the prescribed mean field but not the fluctuation and correlation properties of the pulse. We therefore extend the representation to displaced squeezed continuum states, in which the coherent displacement specifies the mean field and covariance kernels describe the additional quantum correlations. We then examine mode-resolved diagnostics of these correlations and their influence on matter observables.

Recent theory and experiments have established bright squeezed vacuum as a
strong-field driver for charged-particle dynamics, high-harmonic generation,
nanotip photoemission, and atomic tunneling ionization
\cite{gorlach2023_quantum_light,tzur2024_motion,tzur2024_squeezed_hhg,
rasputnyi2024_bsv_hhg,heimerl2025_quantum_light,
jiang2026_bsv_tunneling}; its femtosecond single-shot temporal structure has
also been retrieved by spectral interferometry \cite{kern2026_bsv}.

Such nonclassical fluctuations can be included without changing the prescribed mean field by using a displaced squeezed continuum state \cite{fabre_treps},
\begin{equation}
\ket{\{\alpha\},\zeta}
=
\hat D[\alpha]\hat S[\zeta]\ket{0},
\label{eq:ds_cont}
\end{equation}
where \(\hat D[\alpha]\) is the displacement operator introduced above and
\begin{equation}
\hat S[\zeta]
=
\exp\!\left[
\frac12
\int_0^\infty d\omega
\int_0^\infty d\omega'\,
\left(
\zeta^*(\omega,\omega')\hat a(\omega)\hat a(\omega')
-
\zeta(\omega,\omega')\hat a^\dagger(\omega)\hat a^\dagger(\omega')
\right)
\right].
\label{eq:S_cont}
\end{equation}
The symmetric kernel \(\zeta(\omega,\omega')\) determines which spectral components are quantum correlated by the squeezing. We assume that the kernel is sufficiently
regular, for example square integrable over the relevant spectral interval, so that the squeezed state is mathematically well defined and has a finite mean photon number.

With the ordering in Eq.~\eref{eq:ds_cont}, the mean field is determined by the
displacement:
\begin{equation}
\langle \hat a(\omega)\rangle=\alpha(\omega).
\end{equation}
Thus the matching condition to a prescribed classical pulse is unchanged. With
\(\Delta\hat a(\omega)=\hat a(\omega)-\alpha(\omega)\), squeezing instead
changes the normal and anomalous covariances
\begin{equation}
\begin{aligned}
\mathcal N(\omega,\omega')&=
\langle\Delta\hat a^\dagger(\omega)\Delta\hat a(\omega')\rangle,\\
\mathcal M(\omega,\omega')&=
\langle\Delta\hat a(\omega)\Delta\hat a(\omega')\rangle .
\end{aligned}
\label{eq:normal_covariance}
\end{equation}
They determine the noise of temporal and spectral quadratures and, through the
field correlation function, fluctuation-sensitive matter observables. Pulses
with the same mean drive can therefore produce, for example, different electron
wave packet widths.

Unlike a coherent state, a squeezed state also contains photons associated with
its covariance:
\begin{align}
\bar N_{\rm tot}&=\int_0^\infty d\omega\,
\left[|\alpha(\omega)|^2+\mathcal N(\omega,\omega)\right],
\nonumber\\
U_{\rm tot}&=\int_0^\infty d\omega\,\hbar\omega
\left[|\alpha(\omega)|^2+\mathcal N(\omega,\omega)\right].
\label{eq:squeezed_total_energy}
\end{align}
Consequently, the coherent energy in Eq.~\eref{eq:energy_alpha} is only the part
associated with the mean waveform. If that waveform is fixed in absolute
amplitude, squeezing adds to the total energy; if only its shape and the measured
total energy are fixed, the displacement must be reduced so that both terms in
Eq.~\eref{eq:squeezed_total_energy} sum to the measured value.

A useful special case is a separable squeezing kernel,
\begin{equation}
\zeta(\omega,\omega')
=
\zeta_0\,\xi(\omega)\xi(\omega'),
\qquad
\int_0^\infty d\omega\,|\xi(\omega)|^2=1.
\label{eq:zeta_sep}
\end{equation}
The separable kernel is particularly useful for describing mode-engineered bright squeezed-vacuum sources operating in a single or nearly single spectral mode.
Introducing the wave packet operator
\begin{equation}
\hat a_\xi
=
\int_0^\infty d\omega\,\xi^*(\omega)\hat a(\omega),
\end{equation}
one obtains the familiar single-wave packet squeezing operator \cite{Schleich2001}
\begin{equation}
\hat S_\xi(\zeta_0)
=
\exp\!\left[
\frac12
\left(
\zeta_0^*\hat a_\xi^2
-
\zeta_0\hat a_\xi^{\dagger 2}
\right)
\right].
\end{equation}
This shows explicitly that a squeezed coherent pulse may have the same mean
classical waveform as the coherent pulse constructed in
Sec.~\ref{sec:recipe}, while possessing nonclassical fluctuations in the
selected pulse mode.

\subsection{Mode-resolved and broadband intensity correlations}

The difference between states with the same mean field can be measured without
reconstructing the full state. Let \(f_j(\omega)\) define orthonormal spectral wave packet modes selected
by a mode-resolved measurement, and write
\begin{equation}
\hat b_j=\int_0^\infty d\omega\,f_j^*(\omega)\hat a(\omega),
\qquad
\hat n_j=\hat b_j^\dagger\hat b_j .
\end{equation}
Spectral-mode-resolved auto- and cross-correlations are
\begin{equation}
g^{(2)}_{jk}=
\frac{\langle\hat b_j^\dagger\hat b_k^\dagger
\hat b_k\hat b_j\rangle}
{\langle\hat n_j\rangle\langle\hat n_k\rangle}.
\label{eq:g2_resolved}
\end{equation}
This definition is the spectral-mode counterpart of the usual
two-time Glauber correlation $g^{(2)}(t,t')$: the operators
$\hat b_j$ and $\hat b_k$ project the field onto selected spectral
wave packet modes rather than sampling it at two times. Both quantities derive
from the same normally ordered four-point field correlation; the Fourier
transformation acts on the field operators, not directly on the normalized
$g^{(2)}$.
For a detector that instead collects a set \(W\) of orthogonal colors,
\(\hat N_W=\sum_{j\in W}\hat n_j\), and
\begin{equation}
g_W^{(2)}=
\frac{\langle\hat N_W(\hat N_W-1)\rangle}
{\langle\hat N_W\rangle^2}.
\label{eq:g2_multicolour}
\end{equation}
For a coherent pulse, every resolved or integrated value with a nonzero
denominator equals unity, even when \(\alpha(\omega)\) is spectrally structured.
A deviation from unity therefore rules out a coherent state, whereas \(g^{(2)}=1\) by itself does not imply that the field is in a coherent state.

For the separable squeezed mode in Eq.~\eref{eq:zeta_sep}, define
\begin{align}
\zeta_0&=re^{i\theta},\qquad
\alpha_\xi=\int d\omega\,\xi^*(\omega)\alpha(\omega),
\nonumber\\
n_s&=\sinh^2r,\qquad
m_s=-e^{i\theta}\sinh r\cosh r.
\label{eq:squeezed_mode_moments}
\end{align}
Gaussian moment factoring gives
\begin{equation}
g_\xi^{(2)}=
\frac{|\alpha_\xi|^4+4|\alpha_\xi|^2n_s
+2\operatorname{Re}(\alpha_\xi^{*2}m_s)+3n_s^2+n_s}
{(|\alpha_\xi|^2+n_s)^2}.
\label{eq:g2_squeezed_mode}
\end{equation}
For \(r=0\) and \(\alpha_\xi\neq0\), it reduces to unity; for squeezed
vacuum, \(\alpha_\xi=0\), it equals \(3+1/n_s\). For nonzero displacement, the phase-sensitive term distinguishes whether the
quadrature along the displacement is squeezed or antisqueezed. It vanishes for
squeezed vacuum; direct determination of the squeezing angle requires a
phase-referenced quadrature measurement such as balanced homodyne detection.
Equation~\eref{eq:g2_resolved} describes correlations between coherently
selected wave packet modes. Conventional spectrally resolved detection instead
measures frequency-bin photon numbers, which approach this description only
when each channel effectively selects a single relevant mode. Spectral
cross-correlations may be diluted by color integration, so the resolved and
broadband measurements are complementary \cite{glauber,mandel_wolf}.

\subsection{Interaction of a free electron with a continuous--mode field}
\label{sec:application}

To expose directly how the state specification affects matter dynamics, we
consider the exactly solvable one-dimensional free-electron model of
Ref.~\cite{hack_wavepacket_2026}, now with a genuine frequency continuum. The
electron moves along a linear-polarization axis \(x\), has signed charge \(q\)
(\(q=-e\) for an electron), and interacts with the selected beam mode at
\(\mathbf r_0\). We retain a spectral interval
\(\mathcal W\subset(0,\infty)\) containing the pulse. Let
\(G_x(\omega)=\mathbf e_x\!\cdot\!\mathcal G(\mathbf r_0,\omega)\) and
\(A(\omega)=G_x(\omega)/\omega\). A frequency-dependent mode-phase convention
may be chosen such that \(A(\omega)\) is real. In the dipole and
nonrelativistic approximations, after the diamagnetic term has been included in
the normal-mode definition as in Ref.~\cite{hack_wavepacket_2026}, the reduced
Hamiltonian is
\begin{align}
\hat H&=\frac{\hat P^2}{2m}
{}+\int_{\mathcal W} d\omega\,\hbar\omega
\hat a^\dagger(\omega)\hat a(\omega)
{}-\frac{q}{m}\hat P\hat A,
\nonumber\\
\hat A&=\int_{\mathcal W} d\omega\,A(\omega)
\left[\hat a(\omega)+\hat a^\dagger(\omega)\right].
\label{eq:electron_continuum_H}
\end{align}
The coupling is given by
\begin{equation} \kappa(\omega)=\frac{qA(\omega)}{m\hbar\omega},
\end{equation} and we introduce the auxiliary function
\begin{equation} \Gamma(\omega,t) = \kappa(\omega)\bigl(1-e^{-i\omega t}\bigr)
\end{equation} for later convenience.
The exact Heisenberg solution is derived in
\ref{app:electron_dynamics}. Its field-operator-independent part is
denoted by
\(\hat X_{\rm e}(t)=\hat X(0)+\beta_\kappa(t)\hat P\), where
\(\beta_\kappa(t)\) is given in Eq.~\eref{eq:app_electron_beta}.

Assume an initially factorized electron--field state, without requiring
factorization among the field frequencies, and write
\(\langle\hat a(\omega)\rangle=\alpha(\omega)\). With
\begin{equation}
A_{\rm cl}(t)=2\operatorname{Re}\int_{\mathcal W} d\omega\,
A(\omega)\alpha(\omega)e^{-i\omega t},
\end{equation}
the exact solution gives
\begin{equation}
\langle\hat X(t)\rangle
=\langle\hat X_{\rm e}(t)\rangle
-2\hbar\operatorname{Im}\int_{\mathcal W} d\omega\,
\Gamma(\omega,t)\alpha(\omega)
=\langle\hat X_{\rm e}(t)\rangle
-\frac{q}{m}\int_0^t ds\,A_{\rm cl}(s).
\label{eq:electron_mean_position}
\end{equation}
Thus the first field moment fixed by the laboratory matching condition produces
exactly the corresponding semiclassical displacement within this model; no
large-photon-number approximation is required.

The second field moments control the wave packet width. Project the covariance
kernels in Eq.~\eref{eq:normal_covariance} onto the electron response,
\begin{align}
I_\Gamma(t)&=\int_{\mathcal W} d\omega\,|\Gamma(\omega,t)|^2,
\nonumber\\
\mathcal N_\Gamma(t)&=\int_{\mathcal W} d\omega
\int_{\mathcal W} d\omega'\,
\Gamma^*(\omega,t)\mathcal N(\omega,\omega')\Gamma(\omega',t),
\nonumber\\
\mathcal M_\Gamma(t)&=\int_{\mathcal W} d\omega
\int_{\mathcal W} d\omega'\,
\Gamma(\omega,t)\mathcal M(\omega,\omega')\Gamma(\omega',t).
\label{eq:electron_projected_covariances}
\end{align}
Initial electron--field factorization then yields
\begin{equation}
\Delta X^2(t)=\Delta X_{\rm e}^2(t)
{}+\hbar^2\left[I_\Gamma(t)+2\mathcal N_\Gamma(t)
-2\operatorname{Re}\mathcal M_\Gamma(t)\right].
\label{eq:electron_position_variance}
\end{equation}
Here \(\Delta X_{\rm e}^2(t)\) is the variance of
\(\hat X_{\rm e}(t)\). The first term in
brackets is the state-independent commutator contribution. For a coherent
pulse \(\mathcal N=\mathcal M=0\), whereas nonclassical fluctuations enter
through the two projected kernels.

\begin{figure}
\centering
\begin{minipage}{0.45\textwidth}
  \centering
  \includegraphics[width=\linewidth]{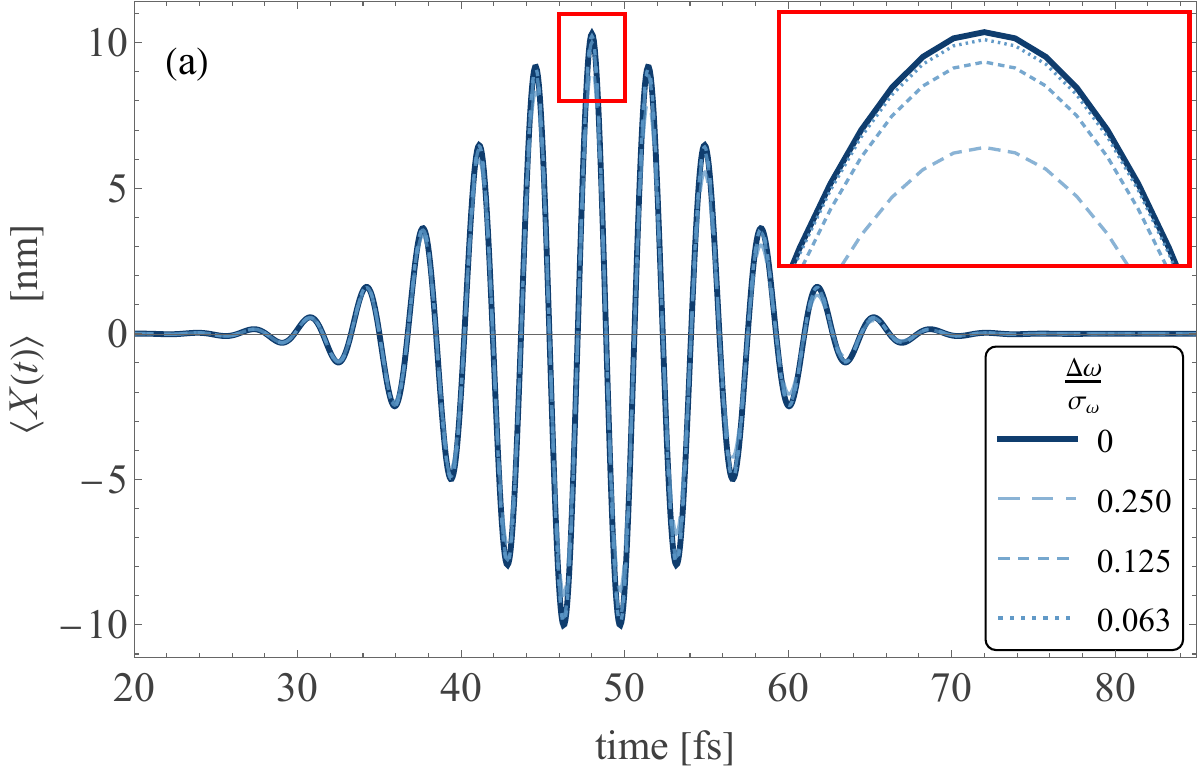}
\end{minipage}
\hfill
\begin{minipage}{0.45\textwidth}
  \centering
  \includegraphics[width=\linewidth]{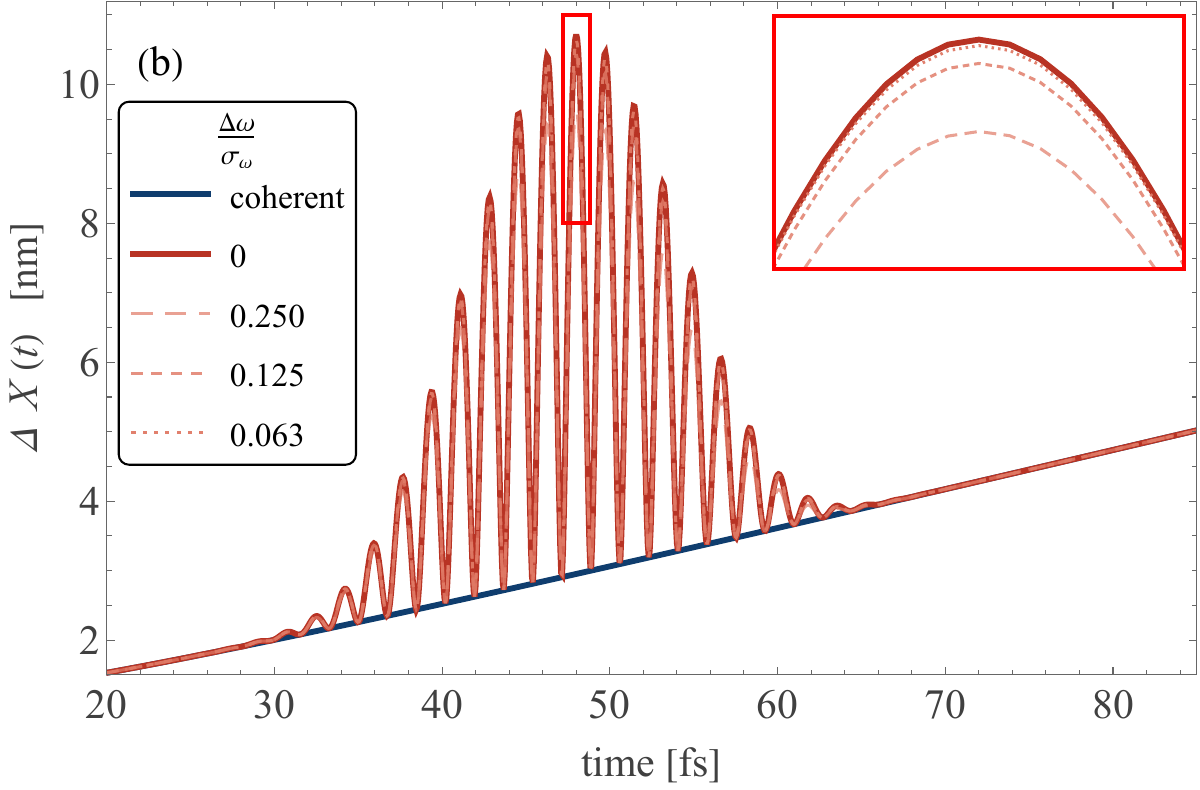}
\end{minipage}
\caption{(a) Mean electron position for the coherent Gaussian pulse specified in the text. The solid curve is the continuum result, while the dashed and dotted curves are obtained from normalized top-hat frequency-bin projections. (b) Position uncertainty for the coherent pulse (blue) and the equal-energy single-wave packet squeezed-vacuum pulse (red). The insets enlarge the marked regions.}
\label{fig:convergence}
\end{figure}

For the separable squeezed mode of Eq.~\eref{eq:zeta_sep}, let
\begin{equation}
K_\xi(t)=\int_{\mathcal W} d\omega\,\Gamma(\omega,t)\xi(\omega).
\label{eq:electron_mode_overlap}
\end{equation}
Equations~\eref{eq:squeezed_mode_moments} and
\eref{eq:electron_projected_covariances} give
\(\mathcal N_\Gamma=n_s|K_\xi|^2\) and
\(\mathcal M_\Gamma=m_sK_\xi^2\), so that
\begin{equation}
\Delta X_{\rm sq}^2(t)-\Delta X_{\rm coh}^2(t)
=2\hbar^2\left[n_s|K_\xi(t)|^2
-\operatorname{Re}\!\left(m_sK_\xi^2(t)\right)\right].
\label{eq:electron_squeezed_excess}
\end{equation}
The overlap \(K_\xi\) makes mode and phase matching explicit: only the squeezed
spectral combination sampled by the electron affects its position noise.

\bigskip

An instructive benchmark compares equal-energy pulses in the same normalized
wave packet mode supported in \(\mathcal W\). Write
\(\xi_c(\omega)=\xi(\omega)e^{i\omega t_c}\), where \(t_c>0\) only places the
finite pulse after the initial time \(t=0\) and changes neither its spectrum nor
its energy. A coherent pulse
\(\alpha_{\rm coh}(\omega)=-i\sqrt{\overline N}\,\xi_c(\omega)\)
and a squeezed-vacuum state
\(\hat S_{\xi_c}(\zeta_0)\ket{0}\) in the same wave packet mode have equal
normally ordered energies exactly when \(n_s=\overline N\); no narrowband
approximation is involved. Set
\(K(t)=K_{\xi_c}(t)\) and choose \(\theta=0\) in this phase convention. With
\begin{equation}
\delta X_{\rm coh}(t)
\equiv\langle\hat X(t)\rangle_{\rm coh}
-\langle\hat X_{\rm e}(t)\rangle
=2\hbar\sqrt{\bar N}\,\operatorname{Re}K(t),
\label{eq:electron_coherent_displacement}
\end{equation}
the exact variance difference is
\begin{equation}
\Delta X_{\rm BSV}^2(t)-\Delta X_{\rm coh}^2(t)
=\delta X_{\rm coh}^2(t)
{}+2\hbar^2\left[\sqrt{\bar N(\bar N+1)}-\bar N\right]
\operatorname{Re}\!\left[K^2(t)\right].
\label{eq:electron_equal_energy_relation}
\end{equation}
The second term is bounded by \(\hbar^2I_\Gamma\) and remains at the vacuum
scale, while the leading term grows as \(\bar N\). Hence, away from zeros of
the coherent excursion,
\begin{equation}
\Delta X_{\rm BSV}^2(t)-\Delta X_{\rm coh}^2(t)
\simeq \delta X_{\rm coh}^2(t)
\qquad (\bar N\gg1).
\label{eq:electron_equal_energy_asymptotic}
\end{equation}
The separable single-wave packet case is an ideal benchmark. A multimode BSV
source must instead be evaluated with the full
\(\mathcal N(\omega,\omega')\) and \(\mathcal M(\omega,\omega')\) kernels in
Eq.~\eref{eq:electron_position_variance}. Bound-state dynamics, ionization,
relativistic motion, depletion, and emission into additional modes lie outside
the present free-electron model.

Fig.~2 illustrates the continuum and frequency-bin results using a shorter pulse than the representative pulse of Table 1, so that the carrier-resolved dynamics and its convergence remain visible. We take \(\lambda_0=1030~\mathrm{nm}\), \(U=0.1~\mathrm{mJ}\), \(w_0=10~\mu\mathrm{m}\), and \(\tau_{\mathrm{FWHM}}=12~\mathrm{fs}\). The pulse is delayed to \(t_c=48~\mathrm{fs}\) through the phase factor in \(\xi_c(\omega)\). At \(t=0\), the electron is described by a minimum-uncertainty Gaussian wave packet with
\(
\langle \hat X(0)\rangle=\langle \hat P\rangle=0,\qquad
\Delta X(0)=1~\mathrm{nm},\qquad
\Delta P=\hbar/{2\Delta X(0)},
\)
and vanishing symmetric position--momentum covariance. The spectral interval is \(\mathcal W=[\omega_0-6\sigma_\omega,\omega_0+6\sigma_\omega]\), with \(\sigma_\omega=\sqrt{2\ln2}/\tau_{\mathrm{FWHM}}\). The displayed discretizations correspond to \(\Delta\omega/\sigma_\omega=0.250,\ 0.125,\) and \(0.0625\).

\section{Relation to other formalisms}
\label{sec:relations}

Continuous-mode quantum optics describes traveling fields by
frequency-resolved operators and coherent amplitudes
\cite{glauber,blow_loudon}. These tools are unchanged here; the distinctive
step is to determine the otherwise abstract displacement \(\alpha(\omega)\)
from the measured analytic signal and pulse energy. This construction is compatible with alternative pulse quantization approaches, such as wave packet mode formalisms or
cavity-inspired discretizations, but keeps the free-space continuum explicit for strong-field applications. With
\(\alpha(\omega)=\alpha_0\xi(\omega)\), the wave packet operator \(\hat a_\xi\)
gives the equivalent single-pulse mode description
\cite{sipe,smith_raymer}. Keeping the continuum explicit is nevertheless
convenient for multimode squeezing or several spectrally resolved detection
channels.

The transverse construction is likewise a mode-basis change and clearly not a
new quantization scheme.
A complete basis built from a Gaussian or an arbitrary normalized seed mode is unitarily equivalent to the plane-wave basis \cite{fabre_treps}. As discussed in Ref. \cite{fabre_treps}, different mode decompositions provide equivalent descriptions of the same field, and the most useful basis is generally the one adapted to the physical problem and the measurement being considered. Apart from the forward-propagating paraxial reduction introduced in Sec.~\ref{sec:paraxial}, the transverse-mode transformation itself is exact. A further physical approximation is made only when orthogonal spatial or polarization modes are discarded; this distinction matters for space--frequency-coupled or polarization-structured pulses.

The normalized top-hat bins of Sec.~\ref{sec:freq_bins} are numerical projections of the continuum and should not be confused with spectral eigenmode decompositions used to diagnose correlations and modal purity \cite{rohde_spectral}. Their spacing sets the numerical spectral resolution, whereas \(T_{\mathrm{box}}=2\pi/\Delta\omega\) is only the associated auxiliary Fourier window.

Finally, the normalization supplies the incident field state, while  Sec. \ref{sec:application} illustrates its use in an exactly solvable
free-electron model. Beyond this benchmark, depletion, emission into additional modes, field--matter entanglement, and nonclassical radiation emerge from dynamics governed by the chosen light--matter Hamiltonian
\cite{foldi,varro,gonoskov_qosf,theidel_qo_hhg,stammer_absence}. No additional
density-of-states factor is needed to normalize the selected incident mode;
such factors enter when summing over unselected final radiation modes.

\section{Conclusion}

We have presented a practical route from a finite free-space laser pulse to a
normalized continuous-mode quantum state. The construction separates
free-space field quantization, the choice of a transverse vector-mode family,
and the specification of the state occupying that family. An arbitrary
normalized spatial--polarization profile may
be used for the selected mode, while the pulse energy, spectral amplitude, and
spectral phase determine the coherent displacement \(\alpha(\omega)\) through
the mode-matching condition. 

An explicit Gaussian-pulse example demonstrates the normalization procedure and its consistency with the classical pulse energy. Regarding numerical implementations, frequency bins enter only as auxiliary normalized projections for numerical work and do not define a physical quantization volume. The derived error
estimate and the transform-limited and chirped tests show that convergence
requires resolving both the spectral amplitude and phase. More generally,
matching the mean field does not uniquely determine the quantum state:
displaced squeezed pulses can retain the same waveform while changing the
field covariance and total energy, and mode-resolved and broadband
\(g^{(2)}\) measurements provide complementary diagnostics of these
correlations.

As an explicit dynamical test, we solved a continuum-mode free-electron model
in the Heisenberg picture. The displacement \(\alpha(\omega)\) reproduces the
semiclassical mean trajectory without a large-photon-number approximation,
whereas projections of the field covariances  determine the
field-induced contribution to the electron position variance.
For coherent and single-wave packet squeezed-vacuum pulses with the same spectrum and energy, the difference between the squeezed vacuum and coherent position variances equals the squared coherent displacement to leading order, with a phase-sensitive correction at the vacuum level. This exactly
solvable benchmark
shows how the normalized input state enters a matter observable while also
making clear that bound-state dynamics, depletion, and generated radiation
require a broader model. The framework therefore provides a direct bridge
between laboratory pulse specifications and quantum-optical strong-field and
attosecond calculations without assigning physical significance to an
auxiliary quantization box.

\appendix

\section{Continuum limit and normalization convention}
\label{app:V}

For completeness we recall the standard box-to-continuum limit used to fix our
normalization convention. In a quantization volume \(V\), the positive-frequency
electric field is
\begin{equation}
\hat{\mathbf E}^{(+)}(\mathbf r,t)
=
i\sum_{\mathbf k,\lambda}
\sqrt{\frac{\hbar\omega_k}{2\varepsilon_0 V}}
\,\bm{\epsilon}_{\mathbf k\lambda}
\,e^{i\mathbf k\cdot \mathbf r}
\hat a_{\mathbf k\lambda}e^{-i\omega_k t},
\end{equation}
where \(\mathbf k\) is the wave vector, \(\omega_k=c|\mathbf k|\),
\(\lambda=1,2\) labels the two transverse polarizations,
\(\bm{\epsilon}_{\mathbf k\lambda}\) is the corresponding unit polarization
vector, and \(\hat a_{\mathbf k\lambda}\) annihilates the associated box mode.
The discrete operators obey
\begin{equation}
[\hat a_{\mathbf k\lambda},\hat a^\dagger_{\mathbf k'\lambda'}]
=
\delta_{\mathbf k,\mathbf k'}\delta_{\lambda\lambda'}.
\end{equation}
The continuum operators are defined by
\begin{equation}
\hat a(\mathbf k,\lambda)
=
\sqrt{\frac{V}{(2\pi)^3}}\,\hat a_{\mathbf k\lambda},
\qquad
\sum_{\mathbf k}\rightarrow \frac{V}{(2\pi)^3}\int d^3k .
\end{equation}
They obey
\begin{equation}
[\hat a(\mathbf k,\lambda),\hat a^\dagger(\mathbf k',\lambda')]
=
\delta_{\lambda\lambda'}\delta^{(3)}(\mathbf k-\mathbf k'),
\end{equation}
and substitution gives Eq.~\eref{eq:E_plane}. The auxiliary volume \(V\)
therefore cancels from the field operator and cannot enter physical predictions.

\section{Mode-space transformations and normalization}
\label{app:mode_transform}

\subsection{Transverse-mode operators}

Starting from
\begin{equation}
[\hat a(\mathbf k_\perp,k_z,\lambda),
 \hat a^\dagger(\mathbf k'_\perp,k'_z,\lambda')]
=
\delta_{\lambda\lambda'}
\delta^{(2)}(\mathbf k_\perp-\mathbf k'_\perp)
\delta(k_z-k'_z),
\label{eq:app_plane_comm_split}
\end{equation}
let \(\tilde u_{m\lambda}(\mathbf k_\perp;k_z)\) be orthonormal,
\begin{equation}
\sum_{\lambda}
\int d^2k_\perp\,
\tilde u^*_{m\lambda}(\mathbf k_\perp;k_z)
\tilde u_{m'\lambda}(\mathbf k_\perp;k_z)
=
\delta_{mm'} .
\label{eq:app_transverse_orth}
\end{equation}
Completeness in the transverse spatial--polarization Hilbert space means
\begin{equation}
\sum_m
\tilde u_{m\lambda}(\mathbf k_\perp;k_z)
\tilde u^*_{m\lambda'}(\mathbf k'_\perp;k_z)
=
\delta_{\lambda\lambda'}
\delta^{(2)}(\mathbf k_\perp-\mathbf k'_\perp).
\label{eq:app_transverse_complete}
\end{equation}
The corresponding transverse-mode annihilation operators are defined by
\begin{equation}
\hat a_m(k_z)
=
\sum_\lambda
\int d^2k_\perp\,
\tilde u^*_{m\lambda}(\mathbf k_\perp;k_z)
\hat a(\mathbf k_\perp,k_z,\lambda).
\label{eq:app_am_definition}
\end{equation}
Equations~\eref{eq:app_transverse_complete} and
\eref{eq:app_am_definition} also give the inverse transformation
\begin{equation}
\hat a(\mathbf k_\perp,k_z,\lambda)
=\sum_m\tilde u_{m\lambda}(\mathbf k_\perp;k_z)\hat a_m(k_z).
\label{eq:app_am_inverse}
\end{equation}
Using Eq.~\eref{eq:app_plane_comm_split}, the projected commutator is
\begin{align}
[\hat a_m(k_z),\hat a^\dagger_{m'}(k'_z)]
&=
\sum_{\lambda\lambda'}\int d^2k_\perp d^2k'_\perp\,
\nonumber\\[-1ex]
&\quad\times
\tilde u^*_{m\lambda}(\mathbf k_\perp;k_z)
\tilde u_{m'\lambda'}(\mathbf k'_\perp;k'_z)
\nonumber\\
&\quad\times
\delta_{\lambda\lambda'}
\delta^{(2)}(\mathbf k_\perp-\mathbf k'_\perp)
\delta(k_z-k'_z)
\nonumber\\
&=\delta(k_z-k'_z)\sum_\lambda\int d^2k_\perp\,
\tilde u^*_{m\lambda}(\mathbf k_\perp;k_z)
\tilde u_{m'\lambda}(\mathbf k_\perp;k'_z)
\nonumber\\
&=\delta_{mm'}\delta(k_z-k'_z).
\label{eq:app_projected_commutator}
\end{align}
In the last line the delta distribution sets \(k'_z=k_z\), after which
Eq.~\eref{eq:app_transverse_orth} applies. Thus no extra normalization factor is
created by the transverse projection.

An arbitrary normalized seed
\(s_\lambda(\mathbf k_\perp;k_z)\) may be set equal to
\(\tilde u_{0\lambda}\). Applying Gram--Schmidt orthogonalization to any spanning family beginning with this seed produces the remaining 
\(\tilde u_m\). Consequently, aberrated,
deformed, frequency-dependent, or polarization-structured pulse profiles are
allowed; named paraxial modes are conveniences rather than requirements.

To make the field normalization explicit, define the real-space vector mode
\begin{equation}
\mathbf v_m(\mathbf r_\perp,z;k_z)
=\frac{1}{2\pi}\sum_\lambda\int d^2k_\perp\,
\tilde u_{m\lambda}(\mathbf k_\perp;k_z)
\bm\epsilon_{\mathbf k\lambda}
e^{i(\mathbf k_\perp\cdot\mathbf r_\perp+k_z z)}.
\label{eq:app_realspace_mode}
\end{equation}
Parseval's identity gives
\(\int d^2r_\perp\,\mathbf v_m^*\!\cdot\!\mathbf v_{m'}
=\delta_{mm'}\). For a forward paraxial beam,
\(\omega_k\simeq ck_z\) for each transverse angular spectrum. Substitution
of Eq.~\eref{eq:app_am_inverse} into Eq.~\eref{eq:E_plane} then gives
\begin{equation}
\hat{\mathbf E}^{(+)}(\mathbf r,t)
=
i\sum_m\int_0^\infty dk_z\,
\sqrt{\frac{\hbar ck_z}{4\pi\varepsilon_0}}\,
\mathbf v_m(\mathbf r_\perp,z;k_z)\hat a_m(k_z)e^{-ick_z t}.
\label{eq:app_field_kz}
\end{equation}

\subsection{Frequency normalization}
\label{app:frequency_normalization}

For a monotonic branch with \(dk_z/d\omega>0\), set
\begin{equation}
\hat a_m(\omega)=
\sqrt{\frac{dk_z}{d\omega}}\,\hat a_m(k_z(\omega)).
\label{eq:app_frequency_operator}
\end{equation}
Using
\begin{equation}
\delta(k_z(\omega)-k_z(\omega'))
=\frac{\delta(\omega-\omega')}{dk_z/d\omega},
\end{equation}
where \(dk_z/d\omega\) is evaluated at the root \(\omega=\omega'\), one
finds
\begin{align}
[\hat a_m(\omega),\hat a_{m'}^\dagger(\omega')]
&=\sqrt{\frac{dk_z}{d\omega}\frac{dk'_z}{d\omega'}}\,
\delta_{mm'}\delta(k_z-k'_z)
\nonumber\\
&=\delta_{mm'}\delta(\omega-\omega').
\label{eq:app_frequency_commutator}
\end{align}

Within the leading-order forward-paraxial reduction used here, \(k_z=\omega/c\) and \(dk_z/d\omega=1/c\). Selecting \(m=0\), changing variables in
Eq.~\eref{eq:app_field_kz}, and using
\(\hat a_0(k_z)=\sqrt{c}\,\hat a(\omega)\) gives
\begin{equation}
\hat{\mathbf E}^{(+)}(\mathbf r,t)
=i\int_0^\infty d\omega\,
\sqrt{\frac{\hbar\omega}{4\pi\varepsilon_0 c}}\,
\mathbf v_0(\mathbf r_\perp,z;\omega/c)
\hat a(\omega)e^{-i\omega t}.
\label{eq:app_field_frequency}
\end{equation}
Writing
\(\mathbf v_0=\mathbf u_\omega/\sqrt{A_{\rm m}}\) recovers
Eq.~\eref{eq:paraxial_G_convention}.
Equations (\ref{eq:app_field_kz})--(\ref{eq:app_field_frequency}) specify the selected transverse mode in a chosen reference plane. Replacing \(\omega_{\mathbf k}\) by \(ck_z\) removes the \(k_\perp^2\)-dependent propagation phase and therefore does not describe diffractive evolution away from that plane. Such propagation requires retaining
\begin{equation}
k_z(\omega,k_\perp)\simeq \frac{\omega}{c}-\frac{c k_\perp^2}{2\omega}.
\end{equation}
This limitation does not affect the normalization or the local field matching performed in the reference plane.

\section{Frequency-bin discretization}
\label{app:frequency_bins}

For nonoverlapping equal-width bins, Eq.~\eref{eq:bin_definitions_main} gives
\begin{align}
[\hat a_n,\hat a_m^\dagger]
&=\frac{1}{\Delta\omega}
\int_{B_n}d\omega\int_{B_m}d\omega'\,
\delta(\omega-\omega')
\nonumber\\
&=\delta_{nm}.
\label{eq:app_bin_commutator}
\end{align}
Each \(\hat a_n\) annihilates the normalized constant top-hat mode in its
interval. Let
\(W=\bigcup_nB_n\), and define
\begin{equation}
\overline\alpha_n=\frac{1}{\Delta\omega}
\int_{B_n}d\omega\,\alpha(\omega),
\qquad
(P_{\Delta\omega}\alpha)(\omega)=\overline\alpha_n
\quad(\omega\in B_n).
\label{eq:app_bin_projection}
\end{equation}
Then \(\alpha_n=\sqrt{\Delta\omega}\,\overline\alpha_n\) and
\(N_{\Delta\omega}=\int_Wd\omega\,
|P_{\Delta\omega}\alpha|^2\). Orthogonality of this projection gives
the exact identity
\begin{align}
\bar N-N_{\Delta\omega}
&=\int_{\omega\notin W}d\omega\,|\alpha(\omega)|^2
\nonumber\\
&\quad+\sum_n\int_{B_n}d\omega\,
|\alpha(\omega)-\overline\alpha_n|^2.
\label{eq:app_bin_error_identity}
\end{align}
The first term is spectral truncation and the second is unresolved intrabin
structure. Consequently, \(N_{\Delta\omega}\leq\bar N\); the difference is
carried by modes orthogonal to the retained top hats and is not a physical
photon loss. With
\(\bar\omega_n\) at the bin midpoint, Eq.~\eref{eq:bin_invariants_main} also
satisfies
\(U_{\Delta\omega}=\int_Wd\omega\,\hbar\omega
|P_{\Delta\omega}\alpha(\omega)|^2\). The relative photon-number and energy
errors need not coincide for an asymmetric spectrum.

For a spectrum with a square-integrable first derivative, the Poincar\'e
inequality bounds the projection error by
\begin{equation}
0\leq
\sum_n\int_{B_n}d\omega\,
|\alpha-\overline\alpha_n|^2
\leq
\frac{(\Delta\omega)^2}{\pi^2}\int_Wd\omega\,
|\partial_\omega\alpha|^2.
\label{eq:app_bin_poincare}
\end{equation}
For a smooth spectrum and bins that tile \(W\), the sharper asymptotic result is
\begin{equation}
\epsilon_N
=\epsilon_{\rm tail}
+\frac{(\Delta\omega)^2}{12\bar N}\int_Wd\omega\,
|\partial_\omega\alpha|^2+O\!\left((\Delta\omega)^4\right),
\label{eq:app_bin_asymptotic}
\end{equation}
where
\(\epsilon_{\rm tail}=\bar N^{-1}\int_{\omega\notin W}|\alpha|^2d\omega\).
Writing \(\alpha=Ae^{i\Phi_\alpha}\), with \(A=|\alpha|\), gives
\begin{equation}
|\partial_\omega\alpha|^2
=(\partial_\omega A)^2
+A^2(\partial_\omega\Phi_\alpha)^2.
\label{eq:app_bin_amp_phase}
\end{equation}
Both spectral amplitude and phase must therefore be resolved. If the tails are
negligible, define
\begin{equation}
R_\alpha=
\frac{\int d\omega\,|\partial_\omega\alpha|^2}
{\int d\omega\,|\alpha|^2}.
\label{eq:app_bin_Ralpha}
\end{equation}
The global condition \(\Delta\omega\sqrt{R_\alpha}\ll1\) avoids over-weighting
the weak spectral tails. For a target error \(\eta\),
Eq.~\eref{eq:app_bin_asymptotic} suggests
\(\Delta\omega\lesssim\sqrt{12\eta/R_\alpha}\); convergence of the required
observables remains the final test.

The same condition applies on the classical and quantum sides. For an exactly
matched scalar mode, Eq.~\eref{eq:mode_matching} implies
\begin{equation}
R_\alpha=
\frac{\displaystyle\int d\omega\,
\left|\partial_\omega[\mathcal E_{\rm cl}(\omega)/G_0(\omega)]\right|^2}
{\displaystyle\int d\omega\,
\left|\mathcal E_{\rm cl}(\omega)/G_0(\omega)\right|^2},
\label{eq:app_classical_bin_criterion}
\end{equation}
where the irrelevant constant phase has cancelled. The frequency dependence of
the selected mode normalization is retained through \(G_0(\omega)\).

Midpoint sampling is the approximation
\begin{equation}
\hat a_n\simeq\sqrt{\Delta\omega}\,\hat a(\bar\omega_n),
\qquad
\alpha_n\simeq\sqrt{\Delta\omega}\,\alpha(\bar\omega_n),
\label{eq:app_bin_sampling}
\end{equation}
and must be distinguished from the exact top-hat integral in
Eq.~\eref{eq:bin_definitions_main}. It must likewise be tested by reducing
\(\Delta\omega\).

Equally spaced frequencies may be viewed as Fourier modes on an auxiliary
interval
\begin{equation}
T_{\rm box}=\frac{2\pi}{\Delta\omega},
\qquad
\phi_n(t)=\frac{e^{-i\omega_n t}}{\sqrt{T_{\rm box}}},
\qquad
\int_0^{T_{\rm box}}dt\,\phi_n^*(t)\phi_m(t)=\delta_{nm}.
\label{eq:app_Tbox}
\end{equation}
The window must contain the temporal waveform of interest, but is neither the
pulse duration nor a cavity round-trip time. The associated
\(cT_{\rm box}\) is only a Fourier bookkeeping length.

For Table~\ref{tab:bin_convergence}, set
\(x=(\omega-\omega_0)/\sigma_\omega\) and use the normalized amplitude
\begin{equation}
\xi_C(x)=(2\pi)^{-1/4}
\exp\!\left[-\frac{x^2}{4}+i\frac{Cx^2}{2}\right].
\label{eq:app_chirped_gaussian}
\end{equation}
Here \(C\) is the dimensionless quadratic spectral-phase (chirp) parameter.
The exact top-hat coefficients are proportional to
\begin{equation}
\left(\frac{\Delta\omega}{\sigma_\omega}\right)^{-1/2}
\int_{x_n}^{x_n+\Delta\omega/\sigma_\omega}dx\,\xi_C(x).
\end{equation}
Summing their squared moduli over
\(-6\leq x\leq6\) gives the errors reported in Table
\ref{tab:bin_convergence}; the omitted Gaussian tail is
\(1-\operatorname{erf}(6/\sqrt2)=1.97\times10^{-9}\). Since
\begin{equation}
\frac{\int dx\,|\partial_x\xi_C|^2}
{\int dx\,|\xi_C|^2}
=\frac14+C^2,
\label{eq:app_gaussian_derivative_norm}
\end{equation}
Eq.~\eref{eq:app_bin_asymptotic} yields Eq.~
\eref{eq:gaussian_bin_asymptotic_main}.

\begin{table}[t]
\centering
\caption{Convergence of the exact top-hat projection for the Gaussian spectral
mode. The error is \(\epsilon_N=1-N_{\Delta\omega}/\bar N\).}
\begin{tabular}{ccc}
\hline
\(\Delta\omega/\sigma_\omega\) & \(\epsilon_N\), \(C=0\) &
\(\epsilon_N\), \(C=1\) \\
\hline
1.000 & \(2.03\times10^{-2}\) & \(9.25\times10^{-2}\) \\
0.500 & \(5.18\times10^{-3}\) & \(2.53\times10^{-2}\) \\
0.250 & \(1.30\times10^{-3}\) & \(6.46\times10^{-3}\) \\
0.125 & \(3.25\times10^{-4}\) & \(1.62\times10^{-3}\) \\
\hline
\end{tabular}
\label{tab:bin_convergence}
\end{table}

Halving \(\Delta\omega\) reduces the fine-grid error by a factor approaching
four, confirming second-order convergence. For \(\eta=10^{-3}\), the
asymptotic estimate requires about \(4.6\) bins per \(\sigma_\omega\) for
\(C=0\) and \(10.2\) for \(C=1\), or approximately 55 and 123 bins across the
\(12\sigma_\omega\) window. The larger chirped error is caused by phase
cancellation within each bin although \(|\xi_C|^2\) is unchanged.

In practice, convergence should be assessed from the reconstructed field and
the physical observables relevant to the calculation; an individual
\(\alpha_n\) is basis dependent.

\section{Derivation of the continuum free-electron dynamics}
\label{app:electron_dynamics}

This appendix derives the results used in Sec.~\ref{sec:application}. We use
the mode-phase convention in which \(A(\omega)\) and hence
\(\kappa(\omega)\) are real; simultaneous rephasing of the mode operators and
state amplitudes leaves all observables unchanged. The required integrals are
assumed finite, and the parameters are taken in the stable regime
\(m_\kappa>0\), where
\begin{align}
\frac{1}{m_\kappa}&=\frac{1}{m}
-2\hbar\int_{\mathcal W}d\omega\,
\omega|\kappa(\omega)|^2,
\nonumber\\
\beta_\kappa(t)&=\frac{t}{m_\kappa}
+2\hbar\int_{\mathcal W}d\omega\,
|\kappa(\omega)|^2\sin(\omega t).
\label{eq:app_electron_beta}
\end{align}
Since \([\hat P,\hat a(\omega)]=0\), the
momentum-dependent displacement
\begin{equation}
\hat U_\kappa=\exp\!\left\{
\hat P\int_{\mathcal W} d\omega\,\kappa(\omega)
\left[\hat a^\dagger(\omega)-\hat a(\omega)\right]\right\}
\label{eq:app_electron_displacement}
\end{equation}
obeys
\(\hat U_\kappa^\dagger\hat a(\omega)\hat U_\kappa
=\hat a(\omega)+\kappa(\omega)\hat P\). Completing the square in
Eq.~\eref{eq:electron_continuum_H} then gives
\begin{equation}
\hat U_\kappa^\dagger\hat H\hat U_\kappa
=\frac{\hat P^2}{2m_\kappa}
+\int_{\mathcal W} d\omega\,\hbar\omega
\hat a^\dagger(\omega)\hat a(\omega),
\label{eq:app_electron_diagonal_H}
\end{equation}
with \(m_\kappa\) defined in Eq.~\eref{eq:app_electron_beta}. Thus
\(m_\kappa\) is the effective
kinetic parameter of this reduced dressed-mode model, not a QED mass
renormalization.

For the Heisenberg solution, \(\hat P\) is conserved and
\begin{equation}
\frac{d}{dt}\hat a(\omega,t)
=-i\omega\hat a(\omega,t)+i\omega\kappa(\omega)\hat P,
\end{equation}
so that
\begin{equation}
\hat a(\omega,t)=\hat a(\omega)e^{-i\omega t}
+\Gamma(\omega,t)\hat P.
\label{eq:app_electron_a_solution}
\end{equation}
Consequently,
\begin{align}
\hat A(t)&=\hat A_0(t)
+2\hat P\int_{\mathcal W} d\omega\,
A(\omega)\kappa(\omega)[1-\cos(\omega t)],
\nonumber\\
\hat A_0(t)&=\int_{\mathcal W} d\omega\,A(\omega)
\left[\hat a(\omega)e^{-i\omega t}
+\hat a^\dagger(\omega)e^{i\omega t}\right].
\label{eq:app_electron_A_solution}
\end{align}
The position equation is
\begin{equation}
\frac{d}{dt}\hat X(t)=\frac{\hat P}{m}-\frac{q}{m}\hat A(t).
\label{eq:app_electron_Xdot}
\end{equation}
Using \(qA(\omega)/(m\omega)=\hbar\kappa(\omega)\) and
\begin{equation}
-\frac{q}{m}\int_0^t ds\,\hat A_0(s)
=-i\hbar\int_{\mathcal W} d\omega\,
\left[\Gamma^*(\omega,t)\hat a^\dagger(\omega)
-\Gamma(\omega,t)\hat a(\omega)\right],
\label{eq:app_electron_A_integral}
\end{equation}
the terms proportional to \(\hat P\) combine according to
\begin{equation}
\frac{t}{m}-2\hbar\int_{\mathcal W} d\omega\,
\omega|\kappa(\omega)|^2
\left[t-\frac{\sin(\omega t)}{\omega}\right]
=\beta_\kappa(t).
\end{equation}
Integration of Eq.~\eref{eq:app_electron_Xdot} therefore gives
\begin{equation}
\hat X(t)=\hat X_{\rm e}(t)
-i\hbar\int_{\mathcal W}d\omega\,
\left[\Gamma^*(\omega,t)\hat a^\dagger(\omega)
-\Gamma(\omega,t)\hat a(\omega)\right].
\label{eq:electron_position_exact}
\end{equation}

For an initially factorized state
\(\hat\rho(0)=\hat\rho_{\rm e}\otimes\hat\rho_{\rm f}\), the electron and field
parts of Eq.~\eref{eq:electron_position_exact} have zero covariance. Their
means satisfy
\begin{equation}
-2\hbar\operatorname{Im}\int_{\mathcal W} d\omega\,
\Gamma(\omega,t)\alpha(\omega)
=-\frac{q}{m}\int_0^t ds\,A_{\rm cl}(s),
\end{equation}
which proves Eq.~\eref{eq:electron_mean_position}. The purely electronic
variance is
\begin{align}
\Delta X_{\rm e}^2(t)
&=\Delta X^2(0)+\beta_\kappa^2(t)\Delta P^2
\nonumber\\
&\quad+\beta_\kappa(t)
\left\langle\left\{\Delta\hat X(0),\Delta\hat P\right\}\right\rangle.
\label{eq:app_electron_variance}
\end{align}

To evaluate the field part, define
\begin{equation}
\hat B_\Gamma(t)=\int_{\mathcal W} d\omega\,
\Gamma(\omega,t)\Delta\hat a(\omega).
\end{equation}
Then
\(\Delta\hat F_\Gamma=i\hbar(\hat B_\Gamma-\hat B_\Gamma^\dagger)\), and the
continuum commutator gives
\begin{equation}
[\hat B_\Gamma,\hat B_\Gamma^\dagger]=I_\Gamma,
\qquad
\langle\hat B_\Gamma^\dagger\hat B_\Gamma\rangle
=\mathcal N_\Gamma,
\qquad
\langle\hat B_\Gamma^2\rangle=\mathcal M_\Gamma.
\end{equation}
It follows immediately that
\begin{equation}
\langle(\Delta\hat F_\Gamma)^2\rangle
=\hbar^2\left(I_\Gamma+2\mathcal N_\Gamma
-2\operatorname{Re}\mathcal M_\Gamma\right),
\end{equation}
which proves Eq.~\eref{eq:electron_position_variance}. In particular, the
\(I_\Gamma\) term survives for a coherent state because it originates from the
canonical commutator rather than from a normally ordered covariance.

For a squeezed vacuum confined to the normalized wave packet \(\xi_c\),
\begin{equation}
\mathcal N(\omega,\omega')=n_s\xi_c^*(\omega)\xi_c(\omega'),
\qquad
\mathcal M(\omega,\omega')=m_s\xi_c(\omega)\xi_c(\omega').
\end{equation}
The double projections therefore reduce to
\(\mathcal N_\Gamma=n_s|K|^2\) and
\(\mathcal M_\Gamma=m_sK^2\). Let
\begin{equation}
\bar\epsilon_\xi=\int_{\mathcal W} d\omega\,\hbar\omega
|\xi_c(\omega)|^2.
\end{equation}
The normally ordered energies of the coherent and squeezed-vacuum pulses are
\(U_{\rm coh}=\bar N\bar\epsilon_\xi\) and
\(U_{\rm BSV}=n_s\bar\epsilon_\xi\), respectively. Equal energy thus implies
\(n_s=\bar N\) exactly. With \(\theta=0\), write
\(m_s=-s\), where \(s=\sqrt{\bar N(\bar N+1)}\). Equation
\eref{eq:electron_squeezed_excess} becomes
\begin{equation}
\Delta X_{\rm BSV}^2-\Delta X_{\rm coh}^2
=2\hbar^2\left[\bar N|K|^2
+s\operatorname{Re}\!\left(K^2\right)\right].
\label{eq:app_electron_equal_energy_intermediate}
\end{equation}
Writing \(K=x+iy\) and using
\(\delta X_{\rm coh}^2=4\hbar^2\bar N x^2\) rearranges this expression into
Eq.~\eref{eq:electron_equal_energy_relation}. Finally,
\begin{equation}
s-\bar N=\frac{1}{\sqrt{1+1/\bar N}+1}<\frac12,
\qquad
|K|^2\leq I_\Gamma,
\label{eq:variance_vacuum_scale_term}
\end{equation}
where the second inequality follows from the Cauchy--Schwarz inequality. The inequalities in Eq.~\eref{eq:variance_vacuum_scale_term} establish the vacuum-scale bound on the correction in Eq.~\eref{eq:electron_equal_energy_relation}.

\section*{References}

\bibliographystyle{iopart-num}
\bibliography{q_pulses}

\end{document}